\documentclass[preprint,12pt]{elsarticle}

 \usepackage{graphicx}

\usepackage{amssymb}
 \usepackage{amsthm}
\usepackage{hyperref}
\usepackage{xcolor}
\definecolor{dark-red}{rgb}{0.4,0.15,0.15}
\definecolor{dark-blue}{rgb}{0.15,0.15,0.4}
\definecolor{medium-blue}{rgb}{0,0,0.5}
\hypersetup{
    colorlinks, linkcolor={dark-red},
    citecolor={dark-blue}, urlcolor={medium-blue}
}

 \biboptions{comma,round,square}

\journal{Physics Letters B}

\begin{document}

\begin{frontmatter}


\title{Evaluation of the Pion-Nucleon Sigma Term from CHAOS Data}


\author[rvt]{J. Stahov\corref{cor1}}
\ead{jugoslav.stahov@untz.ba}
\author[focal]{H. Clement}
\author[focal]{G. J. Wagner}
\cortext[cor1]{Corresponding author}

\address[rvt]{Faculty of Science, University of Tuzla, Bosnia and Herzegovina, Univerzitetska 4, 75000 Tuzla, Bosnia and Herzegovina}
\address[focal]{Physikalisches Institut, Universit\"{a}t T\"ubingen, Auf der Morgenstelle 14, 72076 T\"ubingen, Germany}
\begin{abstract}
We have reanalyzed the $\pi ^{\pm} p$ scattering data at low energy in the
Coulomb-nuclear interference region as measured by the CHAOS group at TRIUMF
with the  aim to determine the pion-nucleon $\sigma$ term. The resulting value
$\sigma=(44\pm 12)$MeV, while in agreement with lattice QCD
calculations and compatible with other recent analyses, is
significantly lower than that from the GWU-TRIUMF analysis of 2002.
\end{abstract}


\end{frontmatter}


\section{Introduction} During the last decade it became known \cite{Bottino02},\cite{Ellis08},\cite{Bottino08},\cite{Ellis09},\cite{Giedt} that the (spin-independent) cross
section for elastic scattering of supersymmetric cold dark matter particles 
on nucleons depends strongly on the value of the pion-nucleon sigma
term  $\sigma_{\pi N}$. 
This is but one example of the role of the sigma term, a concept that was
introduced in chiral perturbation theory (ChPT) to measure the explicit 
breaking of chiral symmetry due to non-zero masses of light quarks 
\cite{Gasser91}.
The sigma term represents the contribution from the finite quark masses to the
mass of the proton. Its value is related to the strange quark content of the
nucleon.  The pion-nucleon sigma term  $\sigma_{\pi N}$  is also
related to the value of the pion-nucleon invariant amplitude at the unphysical
Cheng-Dashen point where 
$s-u=0$, $t=2m_{\pi}^2$ (Here, $s$, $t$, $u$ 
are the Mandelstam variables). 
Consequently, its determination
is mostly attempted through pion-nucleon scattering experiments at low energies
and by spectroscopy of the hadronic shift and width in the ground state of
pionic hydrogen and deuterium \cite{Gotta08}. 

Ellis {\it et al.} \cite{Ellis08} pleaded strongly for an
\textit{experimental} campaign to better
determine the pion nucleon sigma term. Indeed, due to the difficulty of
experiments with low energy pions, which are notoriously plagued by pion decay
and the emerging muon background, the resulting cross sections from different $\pi N$
 scattering experiments often do not agree with each other and the derived
values of $\sigma_{\pi N}$ differ substantially, often by more than the quoted systematic
errors. Ironically, however, the experimental campaign requested by the authors
of \cite{Ellis08} already existed and had been published \cite{Denz06} after a strong and
long running effort with the CHAOS detector at TRIUMF \cite{Smith95}, which was a
dedicated detector system developed to cope with the peculiarities of low-energy
pion scattering. The experiment provided differential cross sections for elastic
scattering of positively and negatively charged pions off hydrogen at five
energies between $19.9$ and $43.3$ MeV in fine angular steps ranging from the
Coulomb-nuclear interference region to nearly $180^\circ$. 

What was really missing in ref.\cite{Denz06} was a theoretical analysis leading to a
value of $\sigma_{\pi N}$. The authors of ref.\cite{Denz06} merely
considered the isospin-even forward scattering
amplitudes $Re D^+(T_{\pi})$ obtained directly from $\pi^+p$ and $\pi^-p$ differential cross
sections and extrapolated the values to threshold ($T_{\pi}=0$) using the functional
forms given by the KH80 phase shifts \cite{Koch80} or, alternatively, from the more
recent SAID FA02 phase shifts \cite{Arndt04}. In both cases the threshold values and
the related isospin-even scattering lengths turned out to be smaller than in the
previous analyses. From this observation it was qualitatively concluded that the
CHAOS data favour values of $\sigma_{\pi N}$ that are smaller than recently
claimed \cite{Pavan02}. 

It is the purpose of this paper to complete the analysis
quantitatively and to derive the value of the $\pi N$ sigma term using
the CHAOS cross sections as much as possible.This approach differs
substantially from the method favored by the GWU-TRIUMF
group \cite{Pavan02} who uses the huge $\pi N$ data base of SAID hoping that the errors of
partially contradictory measurements are averaged out. We, instead, prefer to
rely as much as possible on the results taken from the most advanced
low-energy pion
spectrometer to date. Furthermore, by adopting the analysis methods and the
notation  of the Karlsruhe group \cite{Hohler83}, we use constraints that warrant
analyticity and unitarity and we exploit forward dispersions relations
of the $\pi N$-scattering amplitudes which ensures, in a sense,
consistency with the existing $\pi N$ data base. For a criticism of the VPI-GWU methods see
\cite{Hohler90}.

\section{Formalism}
The $\pi N$ $\sigma$ term is defined as a matrix element of the quark mass term
\begin{eqnarray}
\sigma_{\pi N}=\frac{m_u+m_d}{4m}\langle p|\bar{u}u+\bar{d}d|p\rangle,\nonumber
\end{eqnarray}
where $m_u$ and $m_d$  are up and down quark masses, and $m$ is the proton mass. It is
an empirical measure of the chiral asymmetry generated by the up and down quark
masses. The $\sigma$ term may be written in the form
\begin{eqnarray}
\sigma_{\pi N}=\frac{m_u+m_d}{4m}\frac{\langle p|\bar{u}u+\bar{d}d-2\bar{s}s|p\rangle}{1-y},\nonumber
\end{eqnarray}
where the parameter $y$, the strange quark content of the proton, is defined as
\begin{eqnarray}
y=2\frac{\langle p|\bar{s}s|p\rangle}{\langle p |\bar{u}u+\bar{d}{d}|p\rangle}.\nonumber
\end{eqnarray}
In a one loop calculation of  chiral perturbation theory (ChPT) Gasser \textit{et al.} \cite{Gasser82}  obtained
\begin{eqnarray}
\sigma_{\pi N}=\frac{35.5\pm 5}{1-y}~ ¸\mathrm{MeV}.\nonumber
\end{eqnarray}
This relation provides a simple way to calculate the strange quark content of the
proton from the known value of the $\sigma$ term.
The low energy theorem \cite{Brown71}, \cite{Dashen71}
relates the so called {\em experimental} $\pi N$ sigma term $\Sigma$ to the
isospin-even scattering amplitude $D^+$ at the Cheng-Dashen point $\nu=0$, $t=2m_{\pi}^2$
\begin{eqnarray}\label{eq:1}
\Sigma=F_{\pi}^2\bar{D}^+(\nu=0, t=2m_{\pi}^2).
\end{eqnarray}
Here $F_{\pi}$ = 92.4 MeV  is the pion decay constant. The pion
mass is denoted by $m_{\pi}$, the nucleon mass by $m$,  $s$, $u$,
$t$ are Mandelstam variables and $\nu={(s-u)}/{4m}$. 
The amplitude $\bar{D}^+$ is defined in terms of the $\pi N$ invariant
amplitudes $A$ and $B$: $D^+=A^++\nu B^+$.  
$\bar{D}^+$  denotes the $D^+$ amplitude from which the pseudo vector
Born term is subtracted: $\bar{D}^+=D^+-D_{Np\nu}^+$.
Within the framework of ChPT Gasser \textit{et al.} \cite{Gasser91} obtained
\begin{eqnarray}\label{eq:2}
\sigma_{\pi N}=\Sigma - \Delta_{\sigma}-\Delta_R,
\end{eqnarray} where \\
\begin{eqnarray}\label{eq:3}
 \Delta_{\sigma}=(15.2\pm0.4)\mathrm{ MeV},\: \Delta_R=0.35\mathrm{MeV}.
\end{eqnarray}
More recent determinations of $\Delta_R$ will be given below.

Since the Cheng-Dashen point is outside the physical region, one has to
perform an analytic continuation of the $\bar{D}^+$ amplitude to this point. Mandelstam
analyticity, unitarity  and crossing symmetry of the $\pi N$ invariant amplitudes are
very strong constraints when analyzing experimental data or when performing an
analytic continuation of the invariant amplitudes outside the physical region.
The most frequently used method for that purpose is the application of
dispersion relations along different curves in the Mandelstam plane. It is
important to stress that use of dispersion relations as a method of analytic
continuation outside the physical region assumes input from the whole energy
region in the physical $\pi N$  channel. Using results from
phase-shift analyses
in dispersion relations implies the use of a whole body of pion-nucleon
data.
Two different methods, both based on dispersion relations, are used
for the
calculation of the $\bar{D}^+$  amplitude at the Cheng-Dashen
point. In the first approach,
the dispersion curves pass through the Cheng-Dashen point which allows
calculating the amplitude $D$ directly \cite{Koch82,Hadzi05,Hite05}. In the second
approach \cite{Pavan02,Gasser88}, one determines the coefficients in the Taylor
expansion of  $\bar{D}^+$ around the center of the Mandelstam triangle (also known as a
sub-threshold expansion). Invariant amplitudes are real inside the Mandelstam
triangle. The $\bar{D}^+$ amplitude is crossing symmetric and is a function of $\nu ^2$
\begin{eqnarray}
\bar{D}^+(\nu ,t)=\sum_{m,n}\bar{d}^+_{mn}\nu^{2m}t^n.\nonumber
\end{eqnarray}
At the Cheng-Dashen point one has:
\begin{eqnarray}
\bar{D}^+(0,2m_{\pi}^2)=\bar{d}_{00}^++\bar{d}_{01}^+\cdot 2m_{\pi}^2+\ldots\nonumber
\end{eqnarray}
\begin{eqnarray}
\Sigma=F_{\pi}^2\cdot \bar{D}^+(0,2m_{\pi}^2)=F_{\pi}^2\cdot (\bar{d}_{00}^++\bar{d}_{01}^+\cdot 2m_{\pi}^2)+\Delta_d\nonumber 
\end{eqnarray}
with
\begin{eqnarray}
 \Sigma=\Sigma_d+\Delta_d\nonumber
\end{eqnarray}
\begin{eqnarray}\label{eq:4}
\Sigma_d=F_{\pi}^2\cdot (\bar{d}_{00}^++\bar{d}_{01}^+\cdot 2m_{\pi}^2).
\end{eqnarray}

$\Sigma_d$  stands for the leading contribution, linear in $t$. The term $\Delta_d$, a
curvature term  \cite{Gasser91}, describes contributions of higher order in $t$. Calculations show
\cite{Gasser91,Hadzi05} that the curvature term is determined mainly by contributions from the
$t$-channel, and is considered as a known quantity.
The final relation that we use to  calculate $\sigma_{\pi N}$  reads:
\begin{eqnarray}\label{eq:5}
 \sigma_{\pi N}=\Sigma_d+\Delta_d-\Delta_\sigma-\Delta_R.
\end{eqnarray}

 In our calculation we use the
value from \cite{Gasser91} $\Delta_d =11.9$ MeV. With the values from
(\ref{eq:3}) this yields $\Delta=\Delta_d-\Delta_\sigma=(-3.3\pm0.4)$
MeV. Other, recent evaluations obtained $\Delta=(-1.8\pm0.2)$ MeV
\cite{Hofer} and $|\Delta_R|\le 2$ MeV \cite{Bernard} instead. But with regard to
the large final uncertainties (see eq.15 below) their application
would not change the result significantly.

In our approach to extract information from the low energy data, we
use for convenience the Lorentz invariant amplitude $C^+(\nu,t)=A^+(\nu,t)+
\frac{4m^2\nu}{4m^2-t}B^+(\nu,t)$.  It is useful to recall
\cite{Hohler83} that
for  $t=0$:  $\bar{C}^+(\nu, t=0)=\bar{D}^+(\nu, t=0)$ 
 and $\bar{d}_{00}^+=\bar{c}_{00}^+$, $\bar{d}_{01}^+=\bar{c}_{01}^+$.
 
Published values of the $\sigma$ term \cite{Koch82,Gasser88,Olsson00,Pavan02,Hadzi05,Hite05,Alarcon12,Yun12}  
range from about 50 MeV to about 75 MeV. The various results depend on
the method and technique used to extrapolate to the Cheng-Dashen
point and on the data input used. The values at the upper end of this 
range were criticised
as corresponding, in the framework of ChPT, to unrealistically large
values of the strange quark content of the nucleon. Therefore, it is
worthwile to mention  recent work \cite{Alarc12} where it is shown in
the framework of covariant heavy baryon perturbation theory that a
large value of the sigma term does not necessarily entail an
implausibly large strange quark content of the nucleon.

\section{Method}
Whichever of the methods mentioned above is used, data from low energy $\pi^{\pm}p$
scattering assure stable and reliable extrapolations of the $\bar{C}^+$ amplitude
to the Cheng-Dashen point. The goal of the CHAOS \cite{Denz06} experiment was to
obtain high quality data needed for that purpose. In the CHAOS experiment $\pi^{\pm}p$
differential cross sections were measured at low energies and at
particularly small angles in
the Coulomb-nuclear interference region, where the known Coulomb non-spin-flip
amplitude $G_c$ \cite{Koch80,Trom77} interferes with the  nuclear amplitude $G^+$ \cite{Hohler83}. The real
part of forward nuclear amplitude $C^+$  is obtained from the experimental scattering data
using the relation
\begin{eqnarray}
&&Re C^+(q^2,t=0)=\frac{4\pi\sqrt{s}}{m}Re G^+(q^2,t=0)=\nonumber \\ 
&&=\lim_{t\to 0}\frac{4\pi\sqrt{s}}{m}\left[\frac{\frac{d\sigma_{\pi^+p}}{d\Omega}-
\frac{d\sigma_{\pi^-p}}{d\Omega}}{4Re G_C(t)}\right]=\lim_{t\to 0}\Delta^{+}(t),\nonumber
\end{eqnarray}
where $G_C$ is the known Coulomb non-spin-flip amplitude,
$\frac{d\sigma_{\pi^{\pm}p}}{d\Omega}$ are  $\pi ^{\pm}p$ differential
cross sections, and $q^2$  is momentum squared in the center of mass
frame. To adapt our notation to the kinematical variables used in the
experiment, the momentum squared is used as an argument of the invariant amplitude $C^+$.
Measurements of differential cross sections for $\pi ^{\pm}p$ scattering in the Coulomb-nuclear interference region allow a determination of the real part of the
forward amplitude $C^+$ at five energies covered by the CHAOS experiment in the low
energy region ranging from $T_\pi$ = 19.9 MeV to 43.3 MeV.

  In the first step of our analysis the
values $\Delta^+$, derived from the CHAOS data at a given energy, are
fitted to a
polynomial of order $N$  in $t$   and extrapolated to the forward
point $t=0$.
The coefficients $a_n$ in a polynomial fit to  $\Delta^+$  are
determined using a robust convergence test function method  \cite{Piet72} by minimizing
\begin{eqnarray}
\chi^2=\chi_{data}^2+\Phi,\nonumber
\end{eqnarray}
\begin{eqnarray}
\chi^2_{data}=\sum_{i=1}^{N_{D}}\frac{(\Delta_i^+-P_N(t_i))^2}{\sigma_i^2}.\nonumber
\end{eqnarray} 
$N_D$ is the number of data at a given energy,  $\Delta_i^+$ and $\sigma_i$ are
experimental values and errors, respectively,  of 
$\Delta^+$, and $P_N$ is the polynomial of order $N$. 
In the convergence test function method a ''penalty''  $\Phi$ is added to
assure a soft cut-off of higher terms in the polynomial expansion. 
It  is of the form 
$\Phi=\lambda\sum_{n=0}^{N}a_n^2(n+1)^3$, where
$\lambda$ is a smoothing parameter which is determined in the method. 
The  convergence test function method was used
in the Karlsruhe-Helsinki phase shift analysis \cite {Koch80}.  There, the 
invariant amplitudes were fitted by polynomials of order 50 to 60 whereas in
our fit to $\Delta^+$ the order does not exceed $N=5 $ (see section 4). 

In the next step, the obtained values of  $Re C^+(q^2,0)$ are fitted
to a polynomial of second order in $q^2$
\begin{eqnarray}\label{eq:6}
Re C^+(q^2,0)=c_0+c_1 q^2+c_2q^4,
\end{eqnarray}
and extrapolated to the threshold $(q^2=0,t=0)$.  It is important to
make a clear distinction 
between the coefficients in expansions (\ref{eq:4}) and
(\ref{eq:6}). The first two coefficients $c_0$ 
and $c_1$ in (\ref {eq:6}) determine $Re C^+(q^2, t=0)$ and its derivative   $\frac{\partial C^+
  (q^2, t=0)}{\partial q^2}$ at threshold, while $\bar{c}^+_{00}$ ($\bar{d}^+_{00}$ )and
$\bar{c}^+_{01}$ ($\bar{d}^+_{01}$) 
in (\ref{eq:4}) are related to $ \bar{C}^+$ and its forward slope  $\frac{\partial{\bar C^+}
  (\nu=0, t)}{\partial t}$ at the center of Mandelstam triangle ($t=0, \nu=0$).
 The basic idea behind our method is just to connect the coefficients in
eq. (\ref{eq:6}), obtained from the analysis of the CHAOS data, 
to the coefficients in (\ref{eq:4}) needed to calculate the pion-nucleon
$\sigma$ term.
To this aim we exploit the analyticity of the  pion-nucleon invariant
amplitudes  through applications of forward dispersion relations  
for the amplitude $C^+$ and its forward slope \cite {Hohler83}.

Using the partial wave decomposition of $Re C^+$ and the effective
range formula for partial waves  $f_{l\pm}=Re\left(\frac{T_{l\pm}}{q}\right)=q^{2l}(a_{l\pm}+b_{l\pm}q^2)$  close
to threshold, one obtains the following approximation of the $C^+ $
amplitude (formula A.3.60 in ref \cite{Hohler83})
\begin{eqnarray}
\left(1-\frac{t}{4m^2}\right)\frac{Re C^+(q^2,t)}{4\pi (1+x)}&=&a_{0+}^++\frac{1}{2}\left(2a_{1+}^+ +a_{1-}^+ -\frac{a_{0+}^+}{4m^2}\right)\cdot t\nonumber \\
&& + \left(b_{0+}^++2a_{1+}^++a_{1-}^++\frac{a_{0+}^+}{2m\cdot m_{\pi}}\right)\cdot q^2\nonumber \\
&& +\:{higher\:  order\: terms\: in}\: t \:{and}\: q^2.\nonumber
\end{eqnarray}
Here, $a_{l\pm}^+$ are isoscalar  $s$-and $p$-wave  scattering lengths,  $b_{0+}^+$ is the $s$-wave
effective range parameter \cite{Hohler83} and
$x=\frac{m_{\pi}}{m}$. According to its definition the coefficient $c_0$ in expansion  (\ref{eq:6}) is connected with $a_{0+}^+$ by 
\begin{eqnarray}
c_0=4\pi (1+x)a_{0+}^+.\nonumber
\end{eqnarray}
The third term on the right hand side is proportional to the
first derivative of the $C^+$ amplitude with respect to $q^2$  at threshold:
\begin{eqnarray}\label{eq:7}
&&\frac{1}{4\pi(1+x)}\left.\frac{\partial Re C^+(q^2,0)}{\partial q^2}\right|_{q^2=0}=b_{0+}^++2a_{1+}^++a_{1-}^++\frac{a_{0+}^+}{2m\cdot m_{\pi}}\nonumber\\
&& 2a_{1+}^++a_{1-}^+=\frac{c_1}{4\pi(1+x)}-b_{0+}^+-\frac{a_{0+}^+}{2m\cdot m_{\pi}}.
\end{eqnarray}
Taking the derivative of $ReC^+$ with respect to $t$ (forward slope), one obtains:
\begin{eqnarray}\label{eq:8}
\frac{1}{4\pi (1+x)}\left.\frac{\partial ReC^+(0,t)}{\partial t}\right|_{t=0}=\frac{1}{2}\left(2a_{1+}^++a_{1-}^++\frac{a_{0+}^+}{4m^2}\right)\approx \frac{1}{2}(2a_{1+}^++a_{1-}^+).
\end{eqnarray}
From eqs. (\ref{eq:7}) and (\ref{eq:8}) one may derive an expression relating \linebreak $\left.\left(\frac{\partial ReC^+(q^2,t=0)}{\partial q^2}\right)\right|_{q^2=0}$ to the
forward slope $\left.\left(\frac{\partial ReC^+(q^2=0,t)}{\partial t}\right)\right|_{t=0}$:
\begin{eqnarray}
\left.\frac{\partial Re C^+(0,t)}{\partial t}\right|_{t=0}=\frac{c_1}{2}-2\pi(1+x)\left(b_{0+}^+ +\frac{a_{0+}^+}{2m\cdot m_{\pi}}\right).\nonumber
\end{eqnarray}
The corresponding forward slope of the amplitude $\bar{C}^+$  is obtained by subtracting the pseudovector Born term:
\begin{eqnarray}
&&\left.\frac{\partial Re \bar{C}^+(0,t)}{\partial t}\right|_{t=0}=\left.\frac{\partial Re C^+(0,t)}{\partial t}\right|_{t=0}-\left.\frac{\partial  C_{Np\nu}^+(0,t)}{\partial t}\right|_{t=0}.\nonumber
\end{eqnarray}
To get the final expression for the coefficient $\bar{c}_{01}^+$, we switch
to the Mandelstam variable $\nu$ as an argument of the $C^+$ amplitude.
In order to determine the coefficient $\bar{d}_{01}^+=\bar{c}_{01}^+=\left.\left(\frac{\partial \bar{C}^+(\nu=0,t)}{\partial t}\right)\right|_{t=0}$ one adds and subtracts the forward slope
at threshold
\begin{eqnarray}
\left.\frac{\partial Re\bar{C}^+(0, t)}{\partial t}\right|_{t=0}
&=&\left.\frac{\partial Re{C}^+(\nu_{th}, t)}{\partial t}\right|_{t=0} -
\left.\frac{\partial {C}^+_{Npv}(\nu_{th}, t)}{\partial t}\right|_{t=0}\nonumber\\
&&+ \left(\left.\frac{\partial Re\bar{C}^+(0, t)}{\partial t}\right|_{t=0} -
\left.\frac{\partial Re\bar{C}^+(\nu_{th}, t)}{\partial t}\right|_{t=0}\right)\nonumber
\end{eqnarray}
\begin{eqnarray}\label{eq:9}
\bar{c}_{01}^+=\frac{c_1}{2}-2\pi(1+x)\left(b_{0+}^++\frac{a_{0+}^+}{2m\cdot m_{\pi}}\right)+\Delta_1,
\end{eqnarray}
where

\begin{eqnarray}
\Delta_1&=&-\left.\frac{\partial C_{Np\nu}^+(\nu_{th},t)}{\partial t}\right|_{t=0}+ \left(\left.\frac{\partial Re\bar{C}^+(0, t)}{\partial t}\right|_{t=0} -
\left.\frac{\partial Re\bar{C}^+(\nu_{th}, t)}{\partial t}\right|_{t=0}\right).\nonumber
\end{eqnarray}

The Born term $C_{Np\nu}^+$ and its derivative are explicitly known (formula A.8.1 in \cite{Hohler83})
\begin{eqnarray}
C_{Npv}^+(\nu, t)=-\frac{g^2}{4m^3}\frac{(m_{\pi}^2-\frac{t}{2})(1-\frac{t}{4m^2})+\nu^2t}{(\nu^2-\nu_B^2)(1-\frac{t}{4m^2})},\nonumber
\end{eqnarray}
\begin{eqnarray}
C_{N}^+(\nu, t)=C_{Npv}^+(\nu, t)-\frac{g^2}{m},\nonumber
\end{eqnarray}

\begin{eqnarray}
\left.\frac{\partial  C_{Np\nu}^+(\nu,t)}{\partial
    t}\right|_{t=0}=-\frac{g^2}{4m^3}\frac{\omega}{\omega^2-\omega_{B}^2}\left(\omega - \frac{m_{\pi}^2}
{\omega +\omega_B}\right).\nonumber
\end{eqnarray}
$\omega$ is the total energy of the pion in the laboratory frame related to variable $\nu$  by $ \nu=\omega+\frac{t}{4m}$, $\nu_{th}=m_\pi+\frac{t}{4m}$, $\omega_B=-\frac{m_{\pi}^2}{2m}$,
( for $t=0$: $\nu=\omega$, $\nu_{th}=m_{\pi}$). $C_{N}^+(\nu, t)$  is a pseudoscalar Born term which appears  in dispersion relations and 
 $g$ is the $\pi$N pseudoscalar coupling constant, which is related to
the pseudovector coupling constant $f$ by
$\frac{g^2}{4\pi}$=$\frac{4m^2}{m_\pi^2}f^2$. In our calculations we
use $f^2$=0.076 \cite{SAID}.\\

 From dispersion relations for the
forward slope of $C^+$  \cite{Hohler83} one obtains the following expression:
\begin{eqnarray}\label{eq:10}
\Delta_1&=&-\left.\frac{\partial Re C_{Np\nu}^+(\nu_{th},t)}{\partial t}\right|_{t=0}\nonumber\\
&&-\frac{2m_{\pi}^2}{\pi}\intop_{m_{\pi}}^{\infty}\frac{d\omega'}{\omega'({\omega'}^2-m_{\pi}^2)}\left.\frac{\partial C^+\left (\omega'+\frac{t}{4m},t\right) }{\partial t}\right|_{t=0}\nonumber \\ && 
-\frac{m_{\pi}}{2m\pi}\intop_{m_{\pi}}^{\infty}\frac{d\omega'}{\omega'^2}\frac{2\omega'+m_{\pi}}{(\omega'+m_{\pi})^2}ImC^+(\omega',0)\nonumber\\
\Delta_1&=& - \left.\frac{\partial  C_{Np\nu}^+(\nu_{th},t)}{\partial t}\right|_{t=0}-I_1-I_2,
\end{eqnarray}
where $I_1$ and $I_2$  are short for the first and the second integral, respectively.
Both integrals are fast converging and may be accurately evaluated using results
from existing phase shift analyses or, by virtue of the optical theorem, from
existing data for total cross sections.

Forward dispersion
relations relate the amplitude  $\bar{C}^+(0) $ at the center of
the Mandelstam triangle  and at the threshold  $\bar{C}^+(m_\pi)$ by the dispersion integral \cite{Hohler83}
 (with the variable $t$  omitted as an variable):
\begin{eqnarray}\label{eq:11}
\Delta_2=\bar{C}^+(0)-\bar{C}^+(m_{\pi})=-\frac{2m_{\pi}^2}{\pi}\intop_{m_{\pi}}^{\infty}\frac{Im C^+(\omega')}{\omega'(\omega'^2-m_{\pi}^2)}d\omega',
\end{eqnarray}
which leads to:
\begin{eqnarray}
\bar{C}^+(0)=\bar{C}^+(m_{\pi})+\Delta_2=C^+(m_{\pi})+1.88 f^2+\Delta_2,\nonumber
\end{eqnarray}
and finally:
\begin{eqnarray}\label{eq:12}
\bar{c}_{00}^+=c_0+1.88 f^2+\Delta_2.
\end{eqnarray}
Eqs. (\ref{eq:9}) and  (\ref{eq:12}) establish the required relations between the coefficients in expansions  (\ref{eq:6}) and  (\ref{eq:4}).

\section{Results}
The CHAOS data \cite{Denz06} consist of differential cross sections grouped in two
angular regions. The first region is in the forward scattering
hemisphere and is defined by values of the Mandelstam variable $t\geq-0.01$ GeV$^2$.
The second group, defined by $t<-0.01$ GeV$^2$, comprises all larger
angles. We have performed two separate
analyses of the CHAOS data. In the first analysis we have included data at
forward angles only. Values of  $\Delta^+(t)$ obtained from the experimental data have been
fitted to a polynomial of third order in $t$. In the second analysis, all data
were included and fitted to a polynomial of order five. In such a way two sets
of values of $Re C^+$ were obtained from the CHAOS data and averaged. This way we
deliberately increased the weight of the small angle data, keeping in mind the
design features of the CHAOS detector. The resulting values of $Re C^+$ are nuclear.
Hadronic values were calculated applying electromagnetic corrections according
to the Nordita procedure from ref. \cite{Trom77} and used as an input for
determination of the coefficients $\bar{c}_{00}^+$ and $\bar{c}_{01}^+$. 

Forward dispersion relations (FDR) are an essential part of our extra\-polation of
the $C^+$ amplitude to the threshold.  In the first step, values of $Re C^+$, obtained from
the CHAOS data, combined with the once subtracted dispersion relations for the $C^+$ 
amplitude, are used to calculate its value at threshold. The FDR for the
amplitude $C^+$, subtracted at the threshold, read \cite{Hohler83}:
\begin{eqnarray}\label{eq:13}
C^+(m_{\pi})&=&Re C^+(\omega)-C_N^+(\omega)+C_N^+(m_{\pi})-I^+
\end{eqnarray}
with
\begin {eqnarray}
I^+&=&\frac{2k^2}{\pi}\intop_{m_{\pi}}^{\infty}\frac{Im C^+(\omega')}{(\omega'^2-\omega^2)(\omega'^2-m_{\pi}^2)}\omega'd\omega',\nonumber 
\end{eqnarray}
where $k$ is the pion lab momentum and   $C_N^+$ the
nucleon Born term  introduced above.  For each energy where $Re C^+$ is determined from the data, one
calculates the dispersion integral and,  using equation (\ref{eq:13}),
finds a corresponding subtraction constant  $C^+(m_{\pi})$. 
The  average value of the five subtraction constants that we obtain this way
is  $C^+(m_{\pi})=(-0.139\pm0.019) m_{\pi}^{-1}$.
We stress that the obtained average value is $not$ our final result
for $c_0$ as defined in (\ref{eq:6}). Rather, it is used as part of the input (square in Fig.1) when calculating the final
values of the amplitude and its forward slope at threshold. 
Partial wave analyses do not give errors of partial
waves so that the error of the integral in (\ref{eq:13}) may not be given and errors
quoted in our results are due only to the numerical procedure. Our estimates
show, however, that errors of the subtraction constant due to the uncertainty of
dispersion integrals are negligible compared to the experimental errors.

The dispersion integrals in (\ref{eq:10}),  (\ref{eq:11}),  and
(\ref{eq:13}) may be accurately evaluated using
available scattering data. In the energy region where results from
phase shift analyses exist ($k\le k_{max}$)
imaginary parts are calculated from partial waves
(GWU/VPI: $k_{max}$ = 2.6 GeV/c; Ka84: $k_{max}$ =6.0 GeV/c).  Tables of total $\pi^{\pm} p$ cross sections, needed to
calculate $Im C^+$, are available up to lab momenta of $k$ = 340 GeV/c \cite{PDG06}.
Parametrization of the total cross sections at high energies are also available
\cite{Don92}. The integrals in (\ref{eq:11}) and  (\ref{eq:13})  are fast converging. More than $97\%$ of their values are
due to contributions below $k$ = 2.6 GeV/c, so that uncertainties from the high
energy parts of the integrals may be neglected.  Using the GWU/VPI solution Fa08
\cite{SAID} one obtains a value of $\Delta_2= - 1.381
m_{\pi}^{-1}$.  $\Delta_1$ has been calculated using the same input.
Our evaluations, using several solutions (Ka84, Fa02, GW06, Fa08), show that the integral $I_1$ is saturated
already  at $\omega_{max}$ which corresponds to the highest lab momentum in the GWU/VPI partial wave solution.
Using the discrepancy method  we found that contributions from higher energies 
to the integral $I_1$ do not exceed  one percent of its value.  Hence,
uncertainties from high energies may be neglected. Using  results from the
GWU/VPI solution Fa08 and tables for total cross sections at high energies,
a value of  $\Delta_1$ = -133.8 GeV$^{-3}$ = -0.364 $m_{\pi}^{-3}$ was obtained.

\begin{figure}[!ht]
\begin{center}
 \includegraphics[width=0.5\columnwidth]{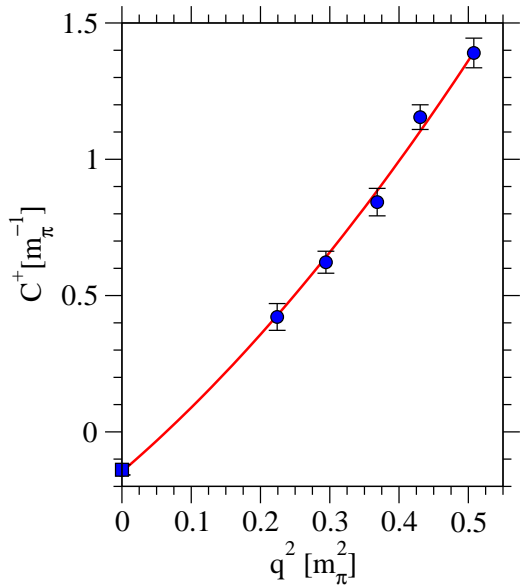}
\caption{\footnotesize Data points: Real parts of the forward $C^+$
  amplitude from the CHAOS experiment \cite{Denz06} (solid circles)
  and their average subtraction constant $C^+(m_\pi)$ (solid square) as described in
  the text. Solid curve: Best fit parabola to the six points.}
\label{Fig.1}
\end{center}
\end{figure}

Fitting a parabola to the five data points for $ReC^+$  
and the average subtraction constant $C^+(m_\pi)$, obtained as described above,
yields the first two coefficients in the expansion (\ref{eq:6}):
\begin{eqnarray}\label{eq:14}
c_0=(-0.140\pm0.013) m_{\pi}^{-1},\nonumber\\
c_1=(2.146\pm0.187) m_{\pi}^{-3},
\end{eqnarray}
which represent essential steps in our analysis. As Fig.1 suggests both coefficients $c_0$ and $c_1$ are well
defined as a result of the small  errors of  $ReC^+$.

The only parameter left to be determined is the  $s$- wave effective range
parameter $b_{0+}^+$. One may follow the authors of ref. \cite{Eric02} and use the Karlsruhe
value  $b_{0+}^+=(-44\pm 7)\cdot10^{-3} m_{\pi}^{-3}$ from ref. \cite{Hohler83}. The very same value was obtained from partial wave
relations derived from the fixed $t$ dispersion relations \cite{Koch86}. Due to the
fact that partial waves from partial wave relations are strictly consistent with
analyticity, the method allows a reliable determination of threshold parameters.
Unfortunately, like some other programs from the Karlsruhe group, Koch's program
for evaluation of partial wave relations was lost. To our knowledge there are no
recent determinations of the s-wave effective range parameters with such a degree
of sophistication. In \cite{Ditsche12} a system of Roy-Steiner equations for ${\pi}N$ scattering was derived. Numerical evaluation of these equations would greatly improve our knowledge about the ${\pi}N$ low energy parameters.

Until results from a new evaluation of partial wave relations become available,
we rely on a simple extrapolation of the $s$-wave to
threshold. Applying the effective range approximation and using the GWU/VPI partial wave solution  FA08  up to $k=80.0$  MeV/c, we obtain $b_{0+}^+=(-52\pm 4)\cdot10^{-3} m_{\pi}^{-3}$.  The value $ b_{0+}^+=(-50\pm 4)\cdot10^{-3} m_{\pi}^{-3}$, which we use
in the present analysis, is a weighted average of the Karlsruhe value and a value
obtained from the FA08 partial wave solution.

Inserting the values of $c_0$ and $\Delta_2$ into (\ref{eq:12}), we
obtain $\bar{c}_{00}^+ = (-1.378 \pm 0.013) m_{\pi}^{-1}$.  The values
of  $\Delta_1$ and $b_{0+}^+$ inserted into  (\ref{eq:9}) lead to
 $\bar{c}_{01}^+ = (1.075\pm 0.098)
m_{\pi}^{-3}$. With these results for $\bar{c}_{00}^+=\bar{d}_{00}^+$  and
$\bar{c}_{01}^+=\bar{d}_{01}^+$ eq.  (\ref{eq:4}) yields $\Sigma_d = (47.2\pm12.2)$ MeV.
Adding the curvature term, we obtain our final result for the $\pi N$ $\sigma$-term as derived
from the CHAOS data: 
\begin{eqnarray}\label{eq:15}
\Sigma= (59\pm 12)\;\mathrm{ MeV} \quad
\sigma_{\pi N} = (44\pm 12)\;\mathrm{ MeV}.
\end{eqnarray}
As stated before, the quoted errors are mainly due to errors having their origin
in the experimental data. Compared to these, errors due to uncertainties of dispersion integrals,
as shown by our calculations, are negligible.

\section{Discussion}
The result obtained for the coefficient $c_0$   corresponds to a $s$-wave scattering
length $a_{0+}^+=(-9.7\pm 0.9)\cdot 10^{-3}m_{\pi}^{-1}$ which is
comparable to the old Karlsruhe result \cite{Hohler83}.  It differs,
however, significantly (and even in sign) from the results based on
the PSI precision experiments on pionic hydrogen and deuterium. These
yielded values of the scattering length that developped over the years
from $(-0.1_{-2.1}^{+0.9})\cdot10^{-3}m_{\pi}^{-1}$ \cite{Schroeder01} to
$(7.6\pm3.1)\cdot 10^{-3}m_{\pi}^{-1}$ \cite{Baru}. The latter value,
however, should not be compared to our result which was obtained from
data in charged pion channels alone. Instead, applying corrections for
isospin violation as may be obtained from Table 6 of ref. \cite{Baru11},  a
value of $(-1.0\pm0.9)\cdot 10^{-3}m_{\pi}^{-1}$ is suggested
\cite{Kubis13} for comparison with our result.

The situation appears much clearer  for p-waves.
The combination $2a_{1+}^++a_{1-}^+$  has been stable during the last three decades. 
Our result  $2a_{1+}^++a_{1-}^+=(199\pm 14)\cdot10^{-3} m_{\pi}^{-3}$,
obtained from (\ref{eq:7}) 
using our values for coefficient $c_1$, effective range $b_{0+}^+$ and
scattering length $a_{0+}^+$,
is in  good agreement with those obtained in partial wave
analyses performed during the last several years
\cite{SAID,Matsinos06,Matsinos12}, e.g. 
$2a_{1+}^++a_{1-}^+=(203.9\pm 1.9)\cdot10^{-3} m_{\pi}^{-1}$
\cite{Matsinos12}.

In view of the spread among the various isoscalar scattering lengths
mentioned above, we performed two tests to study the sensitivity  of
our determination of $\sigma_{\pi N}$ to a variation of the scattering
length. By the first test we demonstrate that it is a frequent, but naive misconception
that a larger scattering length will necessarily result in an increased
$\sigma_{\pi N}$. This is only so if one considers the
contribution of $\bar{c}_{00}^+$ to $\sigma_{\pi N}$ alone,
as may be easily concluded from eqs. (4) and (12). 
However, there is also a strong dependence of $\sigma_{\pi N}$  on the
coefficient $\bar{c}_{01}^+$ i.e. on the forward  slope of $\bar{C}^+$
at the center of the Mandelstam triangle, $t=0$, $\nu=0$. 

This is exactly what we observe when we arbitrarily adopt the recommended
\cite{Kubis13} {\em higher} value  
$a_{0+}^+=(-1.0\pm 0.9)\cdot10^{-3} m_{\pi}^{-1}$   
and obtain the  $C^+$ amplitude for $q^2=0$ in Fig.1. 
Combined with the five data points from the CHAOS experiment we obtain
best fit parameters $c_0=(-0.016\pm 0.014)
m_{\pi}^{-1}$ and  $c_1=(1.400\pm0.184) m_{\pi}^{-3}$ in (\ref{eq:6}).
Hence the {\em higher} value of $c_0$ entails a smaller slope
parameter $c_1$ leading eventually to a ridiculously small  
$\sigma_{\pi N}$=$(7\pm11)$ MeV. Needless to say that this procedure is
inconsistent with our use of forward dispersion relations as described
in section 4. Rather, it serves the purpose to show that it would be a
gross oversimplification to suspect the small isoscalar scattering length as
the sole origin of our low sigma-term.
In addition, the then  resulting value of 
$2a_{1+}^++a_{1-}^+=(149\pm13)\cdot10^{-3}m_{\pi}^{-3}$
would be in disagreement with recently obtained values  
\cite{SAID, Matsinos06, Matsinos12}
unless a large negative value $b_{0+}^+\approx -100\cdot10^{-3}m_{\pi}^{-3}$ 
for the $s$-wave effective range was allowed.

In a second test we assume that the low value of our scattering
length results from a problem with the normalisation of the
CHAOS cross sections. Hence, we retain the shape, notably the slope, 
of the $C^+$  curve in Fig. 1, but shift all 
values by the same amount of +0.013 $m_\pi^{-1}$  in order to get the
intercept with the vertical axis at 
$C^+=c_0= (-0.014\pm0.013)m _{\pi}^{-1}$,  which corresponds to the
recommended value \cite{Kubis13} of $a_{0+}^+=(-1.0\pm 0.9)\cdot 10^{-3}m_{\pi}^{-1}$.
As a result we now obtain 
$c_1=(2.146\pm0.187) m_{\pi}^{-3}$ and $\sigma_{\pi N}=(51\pm12) MeV$.

In order to generate the applied upward shift of the CHAOS data points
in Fig. 1  one has to decrease
the differential cross sections for $\pi^+p$ scattering and/or
increase those for $\pi^-p$ cross sections suitably. (This follows from
the definition of $ReC^+$ in terms of $\pi^{\pm}p$ differential cross
sections  in Sect. 3). 
Quantitatively the required
modifications are compatible (within 2 standard deviations) with the
systematic errors quoted in \cite{Denz06}.
Therefore, we note that the errors of the CHAOS experiment do not
allow a determination of the s-wave scattering length. However, it is
very comforting to observe that the value of $\sigma_{\pi N}$
is robust: it is only by half a standard deviation away from our
result in eq.(\ref{eq:15}).

\section{Conclusions}
Our value of the $\sigma$-term $\sigma_{\pi N}= (44\pm12)$ MeV, based on an analysis of the
CHAOS data, is at the lower end of the range of published values. Given the
systematic uncertainties of the CHAOS experiment one could not expect a
precision determination of its value. Using an analysis which respects the
analytic properties of the $\pi N$ amplitudes we were, however, able
to confirm quantitatively the conjecture of Denz \textit{et al.}
\cite{Denz06} that the sigma term is
small, comparable to the canonical value  $\sigma$ = 49 MeV of Koch
\textit{et al.} \cite{Koch80, Koch82} which remarkably is based on
data from the pre-meson-factory era. This stability is partly owed to 
the constraints imposed by our use of dispersion relations.

Our result agrees, within the (combined) $1\sigma$-errors, 
with those from recent determinations in the framework of covariant 
baryon chiral perturbation theory
\cite{Alarcon12}, \cite{Yun12}, \cite{Luis} 
which range from 41 MeV to 59 MeV. Moreover, it is perfectly compatible with recent lattice QCD
calculations \cite{Giedt,Young} which yielded $\sigma_{\pi N}$=
(47$\pm$9) MeV. However, our result is about 20 MeV smaller than that from the
phenomenological extraction by the GWU-TRIUMF group \cite{Pavan02}. 
With no experimental facility available to improve
the $\pi N$ data base in the forseeable future we considered it
important to challenge that frequently adopted large
value. 

\section{Acknowledgements}
The authors want to thank J.M.Alarc\'on, U.-G. Meissner and
W. Weise for useful and stimulating comments.
We are particularly indebted to B. Kubis for illuminating the role of
isospin breaking in the present context.





\bibliographystyle{model1-num-names}
\bibliography{<your-bib-database>}

\begin{thebibliography}{00}

\bibitem{Bottino02} A. Bottino, F. Donato, N. Fornengo, and S. Scobel,
  Astroparticle Phys. 18 (2002) 205.
\bibitem{Ellis08} J. Ellis, K. A. Olive and C. Savage, Phys. Rev. D77
  (2008) 065026.
\bibitem{Bottino08} A. Bottino, F. Donato, N. Fornengo, and S. Scobel,
  Phys. Rev. D78 (2008) 083520.
\bibitem{Ellis09} J. Ellis, K. A. Olive, and P. Sandick, arXiv:0905.0107v1 [hep.ph].
\bibitem{Giedt} J. Giedt, A. W. Thomas, and R. D. Young,
  Phys. Rev. Lett. 103 (2009) 201802.
\bibitem{Gasser91} J. Gasser, H. Leutwyler and M. E. Sainio, Phys. Lett. B253
(1991) 252 and 260.
\bibitem{Gotta08} D. Gotta \textit{et al.}, Lect. Notes in Phys. 745
  (2008) 165.
\bibitem{Denz06} H. Denz \textit{et al.}, Phys. Lett. B633 (2006) 209.
\bibitem{Smith95} G. R. Smith \textit{et al.}, Nucl. Instr. Meth. A357
  (1995) 296.
\bibitem{Koch80} R. Koch, E. Pietarinen, Nucl. Phys. A336 (1980) 331.
\bibitem{Arndt04} R.A. Arndt, J. Briscoe, I. I. Strakovsky, R. L. Workman and M. M. Pavan, Phys. Rev. C69 (2004) 035213; solution fall 2002.
\bibitem{Pavan02} M. Pavan, R. A. Arndt, I. I. Strakovsky and R. L. Workman, $\pi N$-Newsletter 16 (2002) 110.
\bibitem{Gasser82} J.Gasser, H. Leutwyler, Phys. Reports, 87 (1982) 77.
\bibitem{Brown71} L. S. Brown, W. J. Pardee, R. D. Peccei, Phys. Rev. D4 (1971) 2801.
\bibitem{Dashen71}T. P. Cheng, R. F. Dashen, Phys. Rev. Lett. 26 (1971) 594.
\bibitem{Hohler83} G. H\"ohler, Landolt-B\"ornstein, Vol. 9b2,
  Springer, Berlin 1983 .
\bibitem{Hohler90} G. H\"ohler, Nucl. Phys. A508 (1990) 525c.
\bibitem{Hofer} M. Hoferichter, C. Ditsche, B. Kubis, Ulf-G. Meissner, arXiv1211.1485v1 [nucl-th].
\bibitem{Bernard} V.Bernard, N. Kaiser, Ulf-G.Meissner, Phys. Lett. B389 (1996) 144.
\bibitem{Gasser88} J. Gasser, H. Leutwyler, M. P. Locher, M. E. Sainio, Phys. Letters, B213 (1988) 85.
\bibitem{Koch82}  R. Koch, Z. Phys. C15 (1982) 161.
\bibitem{Olsson00} M. G. Olsson, Phys. Letters B482 (2000) 50.
\bibitem{Hadzi05} M. Hadzimehmedovic, H. Osmanovic, J. Stahov, Int. J. Mod. Phys. A20 (2005) 1876.
\bibitem{Hite05} G. E. Hite, W. B. Kaufmann, R. J. Jacob,
  Phys. Rev. C71 (2005) 065201.
\bibitem{Alarcon12} J. M. Alarc\'on, J. M. Camalich, and J. A. Oller,
  Phys. Rev. D85 (2012) 051503(R) and arXiv:1301.3067v2 [hep.ph].
\bibitem{Alarc12} J. M. Alarc\'on, L. S. Geng, J. Martin Camalich, J. A. Oller, arXiv: 1209.2870 [hep.ph].
\bibitem{Trom77} B. Tromborg, S. Waldenstr\o m, I. \O verb\o, Phys. Rev. D15 (1977) 725.
\bibitem{Piet72} E. Pietarinen, Nuovo Cim. 12A (1972) 522 and Nucl. Phys. B49 (1972) 315.
\bibitem{Stahov97} J. Stahov, $\pi N$ Newsletter 13 (1997) 174.
\bibitem{PDG06} W. M.  Yao \textit{et al.} (PDG), Review of Particle Properties, J. Phys. G33 (2006) 1.
\bibitem{Eric02} T. E. O. Ericson, B. Loiseau, A. W. Thomas, Phys. Rev. C66 (2002) 014005.
\bibitem{Koch86} R. Koch, Nucl. Phys. A448 (1986) 707.
\bibitem{Ditsche12}C. Ditsche, M. Hoferichter, B. Kubis, U. -G. Meissner, arXiv:1203.4758 [hep-ph].
\bibitem{Baru} V. Baru, C. Hanhart, M. Hoferichter, B. Kubis, A. Nogga,
  D.R. Phillips, Phys. Lett. B694 (2011) 473.
\bibitem{Baru11} V. Baru, C. Hanhart, M. Hoferichter, B. Kubis,
  A. Nogga, and D. R. Phillips,  Nucl. Phys. A872 (2011) 69.
\bibitem{Don92} A. Donnachie, P. V. Landshoff, Phys. Lett. B296 (1992) 227.
\bibitem{SAID} SAID, http://gwdac.phys.gwu.edu/.
\bibitem{Matsinos06} E. Matsinos, W. S. Woolcock, G. C. Oades, G. Rasche, A. Gashi, Nucl. Phys. A 778 (2006) 95.
\bibitem{Matsinos12} E. Matsinos,  G. Rasche, J. Mod. Phys. 3 (2012) 1369.
\bibitem{Yun12} Yun-Hua Chen, De-Liang Yao, and H. Q. Zheng,
  arXiv:1212.1893v1[hep-ph].
\bibitem{Schroeder01} H.-Ch. Schr\"oder et. al,
  Eur. Phys. J.  C21 (2001) 473.
\bibitem{Kubis13} B. Kubis, priv. communication.
\bibitem{Young} R. D. Young and A. W. Thomas Phys. Rev. D81 (2010)
  014503.
\bibitem{Luis} L. Alvarez-Ruso, T. Ledwig, J. Martin Camalich, and
  M.J. Vicente-Vacas, archiv:1304.0483
 \end{thebibliography}



\end{document}